\documentclass[twoside]
{article}
\usepackage{epsf}
\usepackage{latexsym}
\usepackage{amssymb}
\usepackage{amsthm}
\usepackage{eucal}
\usepackage{amsfonts}
\usepackage{amscd}
\usepackage{csit}
\usepackage{times}

\title{\bf Event Driven Computations for Relational Query Language}

\author{\large      L.Yu.Ismailova \\[1.52mm]
 {\normalsize { \tt larisa$@$jurinfor.ru}\hfil} 
        \and  \large K.E.Zinchenko\\[1.52mm]
 {\normalsize { \tt kz$@$jurinfor.ru}\hfil} 
        \and  {\large L.V.Bourmistrova}\\[1.52mm]
 {\normalsize { \tt blv$@$jurinfor.ru}\hfil}
        \and
        {\ }\\
        \normalsize Vorotnikovskiy per., 7, bld. 4 \\
        \normalsize Dept. for Advanced Computer Studies and
        Information Technologies \\
        \normalsize Institute for Contemporary Education ``JurInfoR-MSU'' \\
        \normalsize Moscow, 103006 Russia
        \date{} }

\institution{}

\newtheorem{exm}{Example}[section]{\rm }{\rm}
\newtheorem{df}{Definition}[section]

\newtheorem{thm}{Theorem}[section]

\begin{document}

\setcounter{page}{43}
\bibliographystyle{alpha}
\hyphenation{da-ta-ba-ses va-ri-ant}

\markboth{Event Driven Computations for Relational Query Language}
{Workshop on Computer Science and Information Technologies CSIT'99,
Moscow, Russia, 1999}

\maketitle


\begin{abstract}
\noindent This paper deals with an extended model of computations
which uses the parameterized families of entities for data objects
and reflects a preliminary outline of this problem. Some topics
are selected out, briefly analyzed and arranged to cover a general
problem. The authors intended more to discuss the particular
topics, their interconnection and computational meaning as a panel
proposal, so that this paper is not yet to be evaluated as a
closed journal paper. To save space all the technical and
implementation features are left for the future paper.

Data object is a schematic entity
and modelled by the partial function. A notion of type is
extended by the variable domains which depend on events
and types. A variable domain is built from the potential
and schematic individuals and generates the valid
families of types depending on a sequence of events.
Each valid type consists of the actual individuals
which are actual relatively the event or script.
In case when a type depends on the script then corresponding
view for data objects is attached, otherwise a snapshot
is generated.
The type thus determined gives an upper range
for typed variables so that the local ranges are event driven
resulting is the families of actual individuals.
An expressive power of the query language is extended
using the extensional and intentional relations.
\smallskip \hrule width 15em  \smallskip
\noindent {\bf Key words:} event driven model,
individual, partial element, state, script, type, variable domain
\end{abstract}


\section*{Introduction}
\addcontentsline{toc}{section}{Introduction}

An event driven models are known as reflecting the flow
of changes in the problem domains.
In this paper a unit called as {\em data object}  is
persistent under events which enforce changes in its state.
Thus, the data object is observed as a process
in a mathematical sense.

A data object captures both the syntax and semantic features to
increase the flexibility of the entire computational model. The
reasons are to distinguish the {\em outer} events which can
parameterize the behavior of the object to the contrast to the
{\em inner} events. The set of events is determined as a script
giving rise to the dynamic features of data objects.

The inner events enable the evolution of the object
or of the sets of objects and can enforce to change its property.

A short outline of logical background is given in Section~\ref{sect:1}.
The taxonomy of actual, potential and virtual individuals
is used to determine the computation principles for a class
of statements. Both the atomic and compound statements are treated.

In Section~\ref{sect:2} the classes of propositional, actual,
possible and virtual concepts are outlined and covered.

Section~\ref{sect:3} includes the main equations for evaluation
principles. The constructs for types and variable domains
are reviewed.

The evaluation of intentional and extensional predicates in
briefly studied in Section~\ref{sect:4}.

Some features of the event driven relational model
are indicated in Section~\ref{sect:5}. The definitions
are given using the formal descriptions which generate
the additional terms. The subclass of descriptions for
relations with associated formulae is called as restrictions.

The extensions for query language are discussed in Section~\ref{sect:6}
which are based on evaluation within the domain structure.
This structure is based on the notion of variable domains.
The computational features of set theoretic operations are observed.
The generalized junction operation is introduced and studied
relatively events and scripts used as extra parameters.

The main computational ideas are according to the domain structures studied
in~\cite{unpublished-Wol:98a}.
 An approach to common types and abstractions generalizes
those in~\cite{CaWe:85} and is closer to~\cite{EhGoSe:91}.
An approach to construe the variable domains, as in~\cite{Sco:80},
appears to be fruitful to bring into a relational model the event sensitivity,
especially when a meaning of `event flows' is used~\cite{Sco:71}.
Some other features for object-oriented extensions for relational
model, but under restricted assumptions, are studied
in~\cite{Beer:90}, \cite{Mano:90}, they are used in this paper
but in a modified form.

The preliminary variant of this draft paper was accented more
to the properties of a relational model~\cite{unpublished-LYuI:98a}.
The architectures, samples and implementations are outlined
in~\cite{LYuIZ:96}. The feasibility of triggering events
is covered in~\cite{LYuIZ:97}.

\section{Logical background}\label{sect:1}

A general aim is to determine the idealized mathematical
entity with a sensitivity to the events which occur
within the computational environment. There is no reason
to restrict consideration to the constant functions, thus
both the functions and their arguments are assumed to be
dependent on the events.

A common and general idea to evaluate the expressions
means the association with a pair
of objects, function $f$ and argument $a$, some object $f(a)$
by applying $\varepsilon$ function to its argument,
which gives the meaning of the function for this argument:
$\varepsilon[f,a]=f(a)$.
Further advance would be achieved supposing that the functions
and their arguments are schematic and are not restricted to
the class of total functions. A strong candidate to such an entity is
the {\em individual}.

Thus, the notion of evaluation is based on the induced notion of
individual. To determine this notion assume that these entities
can be collected into the domain $H$. This is an important
property of the individuals, and we need, indeed, only the
possibility to construe this collection. Hence, this assumption
seems to be week and not restrictive.

Let the domain $H$ be determined {\em before} the constructing
any theory based on the individuals.
Otherwise, the resulting theory could be contradictive
and non-persistent.
The domain $H$ is assumed to be non-empty and, in fact, is the domain
of all the {\em potential} individuals, which are related to
some theory. `Possible', or `potential' means that they are
schematic and would be implemented into the valid sequences
of the actual states under some events.
Any case they are possible relatively, e.g. some existing theory
of objects

\subsection{Interpretation with the individuals}\label{ind:interp}

Now the events are simulated with the elements of some set $I$,
and the particular event is represented by the index $i\in I$.

For fixed set $I$ of some indices $i$ the sets of {\em actual}
individuals are generated by $U_i \subseteq H$ for any $i\in I$.
There is no one-to-one correspondence between $U_i$ and $i$
because the element of $I$ may be additionally structured. The
elements $i\in I$ are the {\em events}, and the truth values of
expression are evaluated relatively the elements of $I$.

This means that to evaluate the expressions for some language
we need at the first stage to fix the set $I$ and the family
$$
U_i \subseteq H \subseteq V,
$$
where $V$ is the set of {\em virtual} individuals.
The truth value of a statement depends on $i\in I$,
and this principle captures the distinct parts of
the entire statement. Let $1$ and $0$ be the constants $true$
and $false$ respectively. The set $2$ is determined by
$$
2\ =\ \{0,1\}
$$
and gives the set of truth values.

\subsection{Atomic statement}\label{a:ind:atom}

Atomic statements are the most elementary units in a language
and should model the desired event sensitivity. This property
is below left to a semantical consideration.

For any statement
 $\Phi$ the function $\|\Phi\|$ means
the evaluation of $\Phi$ relatively the given
interpretation\footnote{The statements are assumed to be the
closed formulae. This means that they do not contain free variables.
Hence, $\Phi$ is closed.},
which is defined on $I$ with the values from $2$. Thus,
$$
\|\Phi\|i\ =\ 1
$$
means that $\Phi$ is true relatively $i$.
Some other explanation has a sense that
`event $i$ enforces $\Phi$'.
Note, that
$f_i$, $f(i)$, and $fi$ are just the notational variants.
The set of all the functions, which are determined on $I$
and range $2$ is denoted by $2^I$, and
$$
\|\Phi\|: I \to 2\ \textrm{and}\ \|\Phi\| \in 2^I
$$
are the notational variants with the meaning that
$$
\|\Phi\|i : 2,\ \textrm{or}\ \|\Phi\|i \in 2
$$
for $i\in I$, i.e. $\|\Phi\|i = true$ or $\|\Phi\|i = false$.

\subsection{Compound statements}\label{a:ind:compound}

For the logical language with the connectives and quantifiers
a value of the expression is to be determined from the
values of its parts. Let the connectives be
$\neg$, $\land$ and $\lor$ and quantifiers be
$\forall$, $\forall(\cdot)$, $\exists$ and $\exists(\cdot)$.
The evaluation of a statement with the connectives is defined as
shown in Figure~\ref{fig-01}:

\begin{figure*}
\[
\begin{array}{rcllll}
\|\neg \Phi\|i       &=&1&\textrm{iff}&\|\Phi\|i = 0 & \qquad (\neg) \\
\|\Phi \land \Psi\|i &=&1&\textrm{iff}&
     \|\Phi\|i=1\ \textrm{and}\ \|\Psi\|i=1 & \qquad (\land)  \\
\|\Phi \lor \Psi\|i &=&1&\textrm{iff}&
     \|\Phi\|i=1\ \textrm{or}\ \|\Psi\|i=1 & \qquad (\lor)
\end{array}
\]
\caption{The evaluation of a statement with the
       connectives \label{fig-01}  }
\end{figure*}

To evaluate the quantifiers the constants $\bar{c}$ for any $c\in V$
are added:
$$
\|\bar{c}\|i = c
$$
All the evaluation shown in Figure~\ref{fig-02} are valid.
\begin{figure*}
\[
\begin{array}{rcllll}
\|\forall x.\Phi(x)\|i &=&1&\textrm{iff}&\|\Phi(\bar{c})\|i = 1
         \textrm{\ for\ all}\ c\in H &\qquad (\forall) \\
\|\forall(\cdot) x.\Phi(x)\|i &=&1&\textrm{iff}&\|\Phi(\bar{a})\|i = 1
         \textrm{\ for\ all}\ a\in U_i &\qquad (\forall(\cdot))\\
\|\exists x.\Phi(x)\|i &=&1&\textrm{iff}&\|\Phi(\bar{c})\|i = 1
         \textrm{\ for\ some}\ c\in H &\qquad (\exists) \\
\|\exists(\cdot) x.\Phi(x)\|i &=&1&\textrm{iff}&\|\Phi(\bar{a})\|i = 1
         \textrm{\ for\ some}\ a\in U_i &\qquad (\exists(\cdot))
\end{array}
\]
\caption{Evaluations for quantifiers \label{fig-02}  }
\vbox to 1ex {\ }
\end{figure*}

\section{Individual concepts}\label{a:indc}\label{sect:2}

The description ${\mathcal I}$ operator is now used
to introduce individual concept. This operator selects out
the individuals.

\subsection{Individualizing}\label{a:indc:ind}

To determine operator of the description the singletons are
to be established.
\begin{df}[Individual]\label{df:15:01}
An individual $c$ is determined by the singleton \{$c$\}
as shown in Figure~\ref{fig-03}:

\begin{figure*}
\[
\begin{array}{rcllll}
\|{\mathcal I}x.\Phi(x)\|i &=&c&\textrm{iff}&
         \{c\}=\{h\in H:\|\Phi(\bar{h})\|i = 1\}
          &\qquad ({\mathcal I}) \\
\|{\mathcal I}(\cdot)x.\Phi(x)\|i &=&c&\textrm{iff}&
         \{c\}=\{a\in U_i:\|\Phi(\bar{a})\|i = 1\}
          &\qquad ({\mathcal I}(\cdot))
\end{array}
\]
\caption{Determination of the individual\label{fig-03}  } \vbox to
1ex {\ }
\end{figure*}

\end{df}
In the Definition~\ref{df:15:01} for selected $i\in I$
the value $\|\Phi(\bar{h})\|i$ for any $h\in H$ gives the
unique $h$. Then (${\mathcal I}$) generates a {\em possible}
value of the description
relatively $i$, and this value is called $c$. This means that
the description is a function from $I$ into $H$:
$$
\|{\mathcal I}x.\Phi(x)\| : I\to H, \qquad \|{\mathcal I}x.\Phi(x)\|:
i\mapsto c,
$$
where $c$ is an element of the singleton as in the definition above.

The principle $({\mathcal I}(\cdot))$ generates the {\em actual} individuals
as follows:
$$
\|{\mathcal I}(\cdot)x.\Phi(x)\| : I\to U_i,
           \qquad \|{\mathcal I}(\cdot)x.\Phi(x)\|: i\mapsto c,
$$
where $c\in U_i$.

\subsection{Building the concepts}\label{a:indc:buil}

Note that the values of the descriptions range the domain $H$
of possible individuals
whenever the values of the terms range the domain $V$
of virtual individuals. Thus, the values of terms range
the functional space $V^I$.
\begin{df}[Concept]\label{df:15:02}
The elements of functional spaces $2^I$, $U_i^I$, $H^I$, and $V^I$
are called the propositional, actual, possible, and
virtual concepts respectively.
\end{df}
The meaning of a concept is that this is the function
which varies
depending on the assignments, or events from $I$
giving rise to the set of values -- not to unique value.
For instance, the concepts for the term $\tau$ and formula $\Phi$
are respectively the following values:
$$
\|\tau\|:I\to V, \qquad \|\Phi\|:I\to 2
$$
Note that $\|\tau\|$ is an {\em intension} of term $\tau$
and $\|\tau\|i$ is an {\em extension} of term $\tau$
relatively $i\in I$.

A semantic principle is that the intension
of the expression is a function of the intensions of its parts.

\section{Outline of data model}\label{sect:3}

A general view to the computational model is to observe
its arbitrary element as a {\em data object}. The most important
features are  neutrality, adequacy and semantical orientation.

\subsection{Computational features}

A {\em neutrality} results from the principles of computations
which are implemented within the host system. The main rules are
as follows:
\[\begin{array}{lcl}
\|[A,B]\| &=&\langle \|A\|,\|B\|\rangle\ \textrm{and}\\
\|A B\|   &=& \varepsilon \circ \langle\|A\|,\|B\| \rangle,
\end{array} \]
where $[\cdot,\cdot]$ and $\langle \cdot,\cdot \rangle$ are the
different variants of pairing operator, and $\varepsilon$ is an
explicit application. Any object is evaluated according these
rules. The first of them determines an evaluation of the finite
sequences and the second rule indicates an application the
function $A$ to the argument $B$. In particular, a function can
indicate the operator from a relational language, and an argument
gives its operands. As follows from these rules, an evaluation is
independent on any assignment, or event.

\begin{figure*}
\epsfxsize=4in
\centerline{\epsfbox{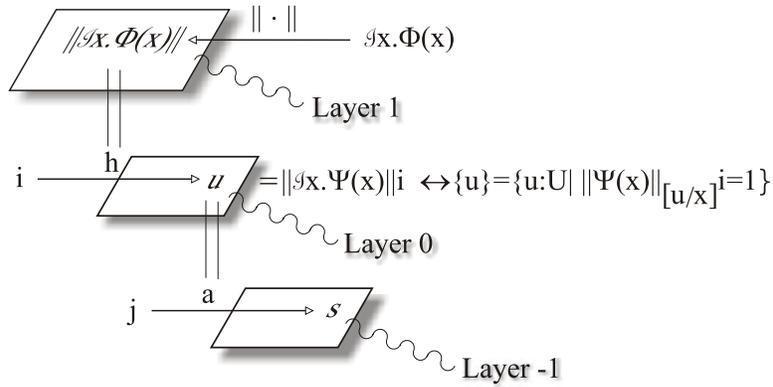}}
\caption{\label{fig:01} {\em Layer 1:}
An individual $h$ is determined by
the evaluated description, so that $h = \|{\mathcal I}x.\Phi(x)\|$.
{\em Layer 0:} An individual $h$ is now a map from events $i$
to actual objects $u$, and $u = h(i)$. On the other hand $u$
is described using the unique indentification principle.
For simplicity, assume that $\Phi(x) = \Psi(x)$.
{\em Layer -1:} An individual $u \equiv a$ is a map from events
to states, and $a(j) = s$.
 }
\vbox to 1ex {\ }
\end{figure*}

An {\em adequacy} takes into account the layers of an evaluation
when the variables range the variable domains. The variables are
understood as driven by the events $i$ from a set of all the
events $Asg$ so that whenever a variable $x$ is of type $T$ then
$\|x\|i \in H_T(\{i\})$ for the variable domain $H_T(\{i\})$ and
the event $i$. When the events are used within the evaluations
then the following rules are valid:
\[\begin{array}{lcl}
\|[A,B]\|i &=&[\|A\|i,\|B\|i]\ \textrm{and}\\
\|A B\|i   &=& (\|A\|i)(\|B\|i),
\end{array} \]
which are similar to their neutral form above. The layers separate
the data objects as shown in Figure~\ref{fig:01}.

The {\em semantics} is based on a definition of the evaluation
map $\| \cdot \|$ while a data object is dropped into the triple
\begin{center}
$\langle$ concept, individual, state$\rangle$
\end{center}

\subsection{Types and variable domains}

A {\em type} is determined by the description
$$
T = {\mathcal I} yi:[A]\forall hi:A (yi(hi)
           \leftrightarrow \|\Phi\|i)
$$
for an arbitrary event $i$, where $\Phi$ is the generator (formula),
$T$ is the type, $h$ is an individual, $A$ is a sort and $[A]$ indicates
the powersort.

A {\em domain} is characterized via the type as follows:
$$
H_T(\{i\}) = \{\hbar | \hbar: \{i\} \mapsto T\},
$$
where $\hbar$ = $hi$. This means that whenever the event is fixed
then the domain has a usual definition.

In case of a {\em variable domain} $H_T(I)$, the definition above
is generalized to the set of the events $I$ so that
$$
H_T(I) = \{h|h:I\to T\},
$$
and this is a family of the usual domains.

The {\em concepts} arise naturally, they depend both on types and
events and are generated as the subsets of variable domains:
$$
C(I) \subseteq H_T(I)
$$

\section{Intensional operations}\label{a:intop}\label{sect:4}

The aims to determine the intentional operations are similar
to those for quantifiers.

\subsection{Intensional atomic predicate}\label{a:intop:iap}

An easiest example is the binary relation
$$
R([ \sigma,\tau])
$$
where $\sigma$ and $\tau$ are arbitrary terms
and $[ \sigma,\tau]$ is an ordered pair.
The awaited meaning of $\|R([ \sigma,\tau])\|$ is a propositional
concept $2^I$, and this is an evaluation of this expression
entirely. The principle of evaluation gives the following:
\[
\begin{array}{rcl}
\|R([ \sigma,\tau])\|i &=& (\|R\|i)(\|[ \sigma,\tau]\|i)      \\
                       &=& (\|R\|i)([\|\sigma\|i,\|\tau\|i])  \\
                       &=& \varepsilon[\|R\|i,[\|\sigma\|i,\|\tau\|i]] \\
          &=& \varepsilon [\|R\|i,<\|\sigma\|,\|\tau\|>i]  \\
          &=& (\varepsilon \circ <\|R\|,<\|\sigma\|,\|\tau\|>>)i
\end{array}
\]
\begin{exm}[Intensional predicate]\label{exm:15:01}
Here we give a verification of type assignment for $\|R([ \sigma,\tau])\|$
and its parts.
\end{exm}
The domains for separate parts of the evaluated expression are
exemplified. To construe the assignment in terms of type--subtype
are tree like derivation is built. The reasons for type assignment
is as follows. The evaluations of terms $\sigma$ and $\tau$ are
$\|\sigma\| : V^I$ and $\|\tau\| : V^I$ respectively. A couple
consisting of $\|\sigma\|$ and $\|\tau\|$ has the type $V^I\times
V^I$, and this is a kind of ordered pair: $<\|\sigma\|,\|\tau\|>$.
Thus, $\|R\|$ has a type from $V^I\times V^I$ into $2^I$. The
derivation below should be read in a direction from bottom to top:
$$
\frac{\hfill \| \sigma\|i : \beta \quad \| \tau\|i : \beta}
     {\displaystyle
\frac{\|R\|i: \alpha \to 2, \qquad \qquad \|[\sigma,\tau]\|i : \alpha \quad}
     {(\|R\|i)(\|[\sigma,\tau]\|i) : 2}
     }
$$
The set of identities for type symbols is as follows:
$$
\beta \equiv V, \quad \alpha \equiv \beta \times \beta
$$
and have the obvious solution $\alpha \equiv V \times V$.
A transition to the concepts results in the derivation:
$$
\frac{\hfill \|\sigma\| : V^I \qquad \| \tau\| : V^I}
     {\|R\|: V^I \times V^I \to 2^I,  \qquad
                 \|[\sigma,\tau]\| : V^I \times V^I \quad}
$$

\subsection{Extensional atomic predicate}\label{a:intop:eap}

This kind of predicates may be exemplified by the usual relations
such that $R \subseteq V \times V$.
\begin{exm}[Extensional predicate]\label{exm:15:02}
An extensional predicate is the constant. Let in a language
this constant be $\overline{R}$ which becomes $R$ when evaluated.
Then
\[
\begin{array}{rcllll}
\|\tau \overline{R} \sigma\|i &=&1&\textrm{iff}&
         [\|\tau\|i,\|\sigma\|i] \in R &\qquad (\overline{R}) \\
\end{array}
\]
\end{exm}





\section{Event driven relational model}\label{sect:5}

The main feature of this model is to support the mutual communications
between the various kinds of objects. The application is based on
the connections of metadata objects, data objects, and states
both within a separate layer and between the distinct layers. Thus,
the establishing and support of the interconnections,
as an actual state, needs some extensible computational environment
with the additional means to declare and manipulate the data objects.

The sublanguage for metadata and data objects has a semantics
which is sensitive to the event changes in accordance with the
consideration given in Section~\ref{a:intop}.

\subsection{Existence of elements}\label{subsect:1-2}

In the discussion above all these objects are driven by the
events, and are partial in their nature. Their behavior needs a
special kind of logic. The logic of partial objects naturally. as
may be shown, relativizes the quantifiers from the greater domains
to the subdomains.

    The everyday mathematical practice gives the valid examples
when the existence of the objects is not well understood. Thus
the identity is supposed to be a trivial relation. Indeed,
`$a=b$' is true when $a$ and $b$ are the same. When they are not
then `$a=b$' is trivially false. All of this is transparent when
both `$a$' and `$b$' are the constant identifiers. If $a$ or $b$
depends on the parameters then their properties are to be
expressed by the equations.

    The more detailed analysis shows that conditionals could be
verified even though there is no complete knowledge of their
possible solutions. The hypothesis or assumption is applied so as
it is valid though its truth value is not determined and it
contains the parameters.

    The notation `${\mathcal E}(\tau )$' is the abbreviation for
`$\tau $ does exist (physically)', contrasting to
`${\exists} (\tau )$' as `$\tau $ can exist (potentially)'.

All the general examples tend to the principle
$$
 {\mathcal E} \tau  \Leftrightarrow {\exists} y.y=\tau ,
$$
where the variable $y$ is unbound in the term $\tau$  and the
expressions could be simplified. Thus, the `does-hold' element
and `can-hold' elements are to be distinguished.

Hence, the predicate of existence $\mathcal E$ is more vital than the
equality and has the priority with respect to the equality.

\subsection{Describing the elements}\label{subsect:4-1}

The existence of partial objects needs a special
mathematical consideration. Generally the question under
discussion arises: is there an object with the predefined
properties and if so then what is its explicit construction? In
any case the predicate {$\mathcal E$} indicates those statements that
mention the actual objects.

    The early discussions of descriptions were mainly informal.
Remark that descriptions fix the individualizing functions. The
purely logical reasons are the following.

    There is no possibility to establish an arbitrary function by
the explicit formula by its instantiations. As in case of
complementary values the instantiations are to be indicated by
some properties. The indirect way to indicate the instantiations
is called `definition by the description' and is denoted by
`${\mathcal I}$'. The descriptions are analogous to quantifiers
and incline to adopt the principle:
\begin{center}
\begin{minipage}{3in}
an entity is equal to the described one iff that entity is
the unique one with the predefined property.
\end{minipage}
\end{center}
To axiomatize the indicated principle of description for any
formula $\phi(x)$ and variable $y$ that is unbound in the
formula the following scheme is assumed:
$$
\forall y\lbrack y = {\mathcal I}x.\phi(x) \Leftrightarrow \forall
x\lbrack \phi(x) \Leftrightarrow x = y\rbrack \rbrack, \eqno ({\mathcal I})
$$
hence, `the described entity is equal to the existent entity'.
Therefore if the described entity is not existent  then
the description ${\mathcal I}x.\phi(x)$ indicates the nonexistent or
indefinite object. The natural (and ambiguous) languages give a
variety of the indefinite entities. Even the rigorous
mathematical languages contain the indefinite objects. Thus the
object ${\mathcal I}x.\neg x=x$ is nonexistent; the object
${\mathcal I}x.x=x$ does exist iff the domain contains at least a single
described element. The sound mathematical ground tends to
generalizations. There are some local universes that contain all
the valid examples.

\subsection{Generating the additional terms}\label{subsect:4-2}

    The descriptions generate the additional terms.
The following theorem covers the general case.

\begin{thm} 
($i$) For arbitrary formula $\Phi(x)$ where the variable $y$ is
unbound the biconditional is valid:
$$
 {\mathcal E}{\mathcal I}x.\Phi(x) \Leftrightarrow {\exists} y \forall
x\lbrack \Phi(x) \Leftrightarrow x=y\rbrack ;
$$
($ii$) ${\mathcal E}{\mathcal I}x.\Phi(x) \Rightarrow
\Phi({\mathcal I}x.\Phi(x)). $
\end{thm}
\begin{proof} The proof is straightforward by principle (${\mathcal I}$)
and the laws for equality and quantifiers having in mind the
biconditional:
$$
{\mathcal E}{\mathcal I}x.
        \Phi(x) \Leftrightarrow {\exists} y.y = {\mathcal I} x.\Phi(x).
$$
\end{proof}
The expanded version of a language when enriched by the principle
of description (${\mathcal I}$) induces the increase of the total term
amount. The additional terms are to be introduced in all the
axioms and rules.

\subsection{Restrictions}\label{subsect:4-3}

    The descriptions are the useful tool. They
maintain the local universes of discourse that are often called
(primitive) frames. Often the researcher is forced to declare a
system of restrictions that are assumed to be independent
objects. By the word let us create all the statements concerning
some formula $\phi$. At the moment do not bother of a special
language of restrictions that is equipped with the diagrams to
represent frames.

One of the possible solutions is given by the relations.
Herein the binary relation $R$ is considered within the
universe generated by formula $\Psi$.

The relation of this kind are schematic.  The descriptions induce
the basic representation of the object and the result is called
the `restriction':
$$
 \tau|\phi \equiv {\mathcal I}x\lbrack x=\tau \&  \phi\rbrack
                                           \eqno ({\bf rest})
$$
for the variable $x$ unbound both in $\tau$ and in $\phi$.

    The intuitive idea is transparent: the object $\tau|\phi$
exist and is equal to $\tau$ while formula $\phi$ is true. To apply it
to the schematic relation $R$, we describe the relation $R$ in a
restricted way:
$$
 R|\Psi  \equiv {\mathcal I}z\lbrack z=R\lbrack x,y\rbrack  \&  \Psi
(x,y)\rbrack.
$$
In fact, $\tau$ exists the less time than the universe object. This
kind of supervision is fruitful in higher order logic when
$\tau|\phi$ could be an element of the class that does not
contain the {\em whole} $\phi$. In particular the relation $R$ above
exists less time than the local universe given by the
formula $\Psi$.
Thus the object $R(x,y)|\Psi (x,y)$ represents the desirable
class.


The restrictions and supplementary universe of discourse are
the valuable tool. All the properties (1)--(13) are understood
within the schemata definitions.
In particular, $R(x,y)|\Psi(x,y)$ is
assumed to be a simple scheme relative (or restricted) by
$\Psi(x,y)$.

\section{Query language}\label{sect:6}

A language for the relational model, or $R$-language,
is just the embedded sublanguage
of the computational model. It depends on the means to identify
the data objects. In case to enable the evaluation of expressions
which are built from metadata, data and states
this language has to have more expressive power than just
a predicate calculus. Thus, the core language is based
on applicative computations and include the operators
of {\em application} and {\em abstraction}:
\[
\begin{array}{rcl}
\varepsilon &:& [\textrm{object},\textrm{object}] \mapsto \textrm{object}, \\
(\lambda\cdot.\cdot) &:& [\textrm{variable}, \textrm{object}] \mapsto \textrm{object}
\end{array}
\]
where $\varepsilon$ takes a pair of objects $f, x$ and results in
$f(x)$, an application of $f$ to $x$ which in turn is the object, and
$(\lambda\cdot.\cdot)$ acts on one variable and one object
resulting in the object.
This definition can be rewritten with the domains $H$ of the partial
elements so that
\[
\begin{array}{rcl}
\varepsilon &:& H \times H \to H, \\
(\lambda\cdot.\cdot) &:& \textrm{variable} \times H \to H
\end{array}
\]
Both the definition and manipulation counterparts of $R$-language
are embedded in some applicative language which is equipped
with the type (sort) system.

For $R$-language the system of {\em types} (attributes)
is built as an inductive class of some metaobjects.
On the other hand, the inductive class of the objects is
assumed as a natural source for $R$-terms and $R$-formulae.
The atomic $R$-formula has a head with a predicate symbol
followed by the objects which are the subject constants.
Both the terms and formulae as typed.

\subsection{Relational structure}\label{subsect:5-1}

The connection between the {\em terms} and data object model
is to be captured by a prescribed prestructure
which assigns to any type symbol $\tau$ the corresponding
domain $H_\tau$. There is a clear reason to select out
the applicative prestructure
$$
(\{H_\sigma\},\{\varepsilon_{\sigma\tau}\}),
$$
where $\sigma$ and $\tau$ are the parameters indicating
the metaobjects (types),
$H_\sigma$ is a family of concepts, $\varepsilon_{\sigma\tau}$
is a family of corresponding applications:
$$
\varepsilon_{\sigma\tau}: H_{\sigma\to\tau} \times H_{\sigma}
                          \to H_{\tau}
$$

The connection between $R$-{\em formulae} and data object model
needs in addition the evaluation map, so that
$$
\langle(\{H_\sigma\},\{\varepsilon_{\sigma\tau}\}),
                                      \|\cdot\|\cdot \rangle
$$
is the {\em structure}, where $\|\cdot\|\cdot$ is the evaluation map,
$$
\|\cdot\|\cdot : Tm \times Asg \to \prod_{\sigma} H_\sigma
$$
where $R$-formulae are from a set $Tm$ of all the terms
(of an applicative system~!) and $Asg$ is a class of the assignments
(events).

The evaluation $\|\cdot\|\cdot$ can be naturally continued
from the class of atomic $R$-formulae to the class of arbitrary
$R$-formulae.

The main features of the computation model are the following:
\begin{description}
\item applicative background,
\item algebraic transparency,
\item natural correspondence of the objects and metaobjects
      via (applicative) prestructure,
\item natural correspondence of the concepts and assignments
      via a structure.
\end{description}

\subsection{Evaluation}\label{subsect:5-2}

The evaluation map has some important particular cases:
\begin{description}
\item the standard relational model is generated in case of
     fixing the of assignments $Asg$. This means an assumption
     that $Asg$ consists of a singular element. Then the
     model properties are derived from
     $$
     \|\cdot\| : Tm \to \prod_\sigma H_\sigma,
     $$
     where $H_\sigma$ is a family of dataobjects, and $Tm$ is
     the set of well defined expressions;
\item the metadata model is derivable from the structure with
     a non-trivial set of assignments $Asg$, possibly, with the
     inner structure.
\end{description}
In fact, the map
\mbox{$\|\cdot\|i : Tm \to \prod_\sigma H_\sigma(\{i\})$} is formally used
where domains $H_\sigma(\{i\})$ correspond to the objects $h(i)\in T$.

The typed data objects are determined by the descriptions.
The {\em type} $T$ corresponds to the formula $\Phi$ as follows:
$$
T = {\mathcal I}y:[A]\forall \hbar : A (y(\hbar)
         \leftrightarrow \Phi)
$$
for any {\em sort} $A$.

Another important join operation, e.g. $\theta$-join, involves
a pair of relations. The counterpart relations are identified by
the formulae $\Phi$ and $\Psi$ respectively, their {\em join}
is determined by
\[\begin{array}{c}
\forall h, \hbar (\theta (h,\hbar) \to \Phi(h)\&\Psi(\hbar)), \\
\forall h (\Phi(h) \to \exists z \forall \hbar (\theta(h,\hbar)
              \leftrightarrow \hbar = z))
\end{array}\]
The often used operations are the set theoretic operations
and join-like operations.

\subsection{Operations}\label{subsect:5-3}

As was noted above, a semantical analysis can be restricted
on the set theoretic and join operations without loss of generality.
The specific features of this analysis is as follows:
\begin{description}
\item the set theoretic operations are the compounds which involve
      the pair of formulae to determine the counterpart relations;
\item in case of join operations, besides evaluating the pair of formulae,
      the interrelations between data objects and the objects
      from a computation environment must be established.
\end{description}

\subsection{Set theoretic operations}\label{subsect:5-4}

\begin{figure*}[t]
$$
({\mathcal J} \circ \langle \|\Phi\| \circ
  \langle J, h\rangle, \|\Psi\| \circ \langle J,\hbar \rangle   \rangle)_f
= {\mathcal J} \circ \langle \|\Phi\|_f \circ \langle J, h \circ
f\rangle,
     \|\Psi\|_f \circ \langle J,\hbar\circ f \rangle \rangle
$$
\caption{The description of dynamics\label{fig-05}  } \vbox to 1ex
{\ }
\end{figure*}
\begin{figure*}[t]
\[
\begin{array}{lcl}
\|\&[(\lambda \theta.)[\overline{h},\overline{\hbar}],
  \&[(\lambda.\Phi)\overline{h},(\lambda.\Psi)\overline{\hbar}]]  \| &=&
{
\& \circ \langle \|\theta\| \circ \langle J,
                                    \langle h, \hbar \rangle \rangle,
  \& \circ \langle \|\Phi\| \circ \langle J,h\rangle,
        \|\Psi\| \circ \langle J, \hbar \rangle \rangle \rangle}
\end{array}
\]
\caption{The equation in a neutral form \label{fig-06}  }%
 \vbox to 1ex {\ }
\end{figure*}
\begin{figure*}[t]
\[
\begin{array}{llcl}
\textrm{atomic object} &\|\theta[\overline{h},\overline{x}]\|i &=&
        (\theta \circ \langle h \circ f, x \circ f \rangle)b \\
\textrm{constant function}
&\|\theta[\overline{h},g(\overline{x})]\|i &=&
        (\theta \circ \langle h \circ f ,
                   g \circ x  \circ f \rangle)b \\
\textrm{ordered pair}
&\|\theta[\overline{h},[\overline{x},\overline{y}]]\|i &=&
        (\theta \circ \langle h \circ f ,
         \langle x \circ f , y \circ f  \rangle \rangle)b  \\
\textrm{application}
&\|\theta[\overline{h},\overline{x}(\overline{y})]\|i &=&
        (\theta \circ \langle h \circ f ,
         \varepsilon \circ \langle x_{f}, y \circ f  \rangle \rangle)b
\end{array}
\]
\caption{The dynamical behavior with evolvent $f : B \to I$
\label{fig-07}  } \vbox to 1ex {\ }
\end{figure*}

The case study of set theoretic operations can be restricted to
the union, intersection and difference of database domains. Let
$\wp$ be the notation for some of the operations listed above.
Then the resulting evaluation in a neutral to the assignments
notation leads to
$$
\|\wp[(\lambda.\Phi)\overline{h},(\lambda.\Psi)\overline{h}] \|
= \wp \circ \langle \|\Phi\|,\|\Psi\|\rangle
      \circ \langle J, h\rangle
$$
This means that the steps of evaluation are as follows:
\begin{description}
\item select out the operation $\wp$,
\item select out the operands $\lambda.\Phi$ and $\lambda.\Psi$,
\item generate the elements $\overline{h}$ from the operands,
\item the effect of the operation $\wp$ is generated by
      selecting and accumulating the evaluated data objects $h$
      which match the operation and operands.
\end{description}
This procedure is almost the same as an evaluation with the
standard data model with the important exception: {\em both the
operation and the operands can be the objects taken from a
computation environment}.

The generalization to the evolvent $f:B \to I$ intends the dynamics
of data objects leading to (in a neutral form):
$$
( \wp \circ \langle \|\Phi\|,\|\Psi\|\rangle
      \circ \langle J, h\rangle)_f
=  \wp \circ \langle \|\Phi\|_f,\|\Psi\|_f\rangle
      \circ \langle J, h \circ f\rangle,
$$
where $f$ in a subscript position indicates the {\em restriction}
so that `the events evolve along the evolvent (script) $f$',
giving the needed transition effect.

Note, that the equation above determines the {\em view} shifting
in accordance to the procedure as follows:
\begin{description}
\item[$a)$] an {\em old} view $i$ is fixed, and $i\in I$;
\item[$b)$] $f$-shifted evaluations of the operands are generated
      resulting in $\|\Phi\|_f$ and $\|\Psi\|_f$;
\item[$c)$] the $f$-shifted data objects $h \circ f$ are accumulated;
\item[$d)$] the relation thus generated is assigned to the {\em new} view
      $b$, and $b\in B$.
\end{description}
Note that the events evolve {\em from} $I$ {\em to} $B$ when the
evolvent $f: B \to I$ is considered. Another observation leads
directly to some important particular case when $f\equiv 1_I$ is
an identity map. In this case the data model gives up the dynamic
behavior becoming the {\em static} model with a trivial script
observing as an identity map.

\subsection{Junction: generalized operation}\label{subsect:5-5}

Some other observation shows the way to bring in more generality
with the set theoretic operations. There is not necessary to
immediately and explicitly indicate the operations.

An {\em implicit} indication of the operation as an attached
procedure binds the selected out data objects with some rule. In
this case the evaluated expression matches the equation below:
$$
\|{\mathcal J}[(\lambda.\Phi)\overline{h},(\lambda.\Psi)\overline{\hbar}]\|
= {\mathcal J} \circ \langle \|\Phi\| \circ
    \langle J, h\rangle, \|\Psi\| \circ \langle J,\hbar \rangle   \rangle,
$$
where $\mathcal J$ is called the `junction' operation.

The evaluation procedure is shown in
Figures~\ref{fig-05}--\ref{fig-07} and is similar to those for the
set theoretic operations $\wp$ excepting the last step which has
to be modified as follows:
\begin{description}
\item[$d')$] a result of the junction $\mathcal J$ operation is generated
by selecting out and accumulating the evaluated data objects
$h$ and $\hbar$ which {\em separately} match both the operation
and operands.
\end{description}
The dynamical behavior is determined by the equation shown in
Figure~\ref{fig-05} for the evolvent $f: B \to I$ in a neutral
formulation. In case of $f\equiv 1_I : I \to I$ for identity map
the {\em static} evaluation is considered.

An evaluation of junction has an important application when the
{\em join}-like operations are used. In this case the expression
of a language contains a conjunction $\&$ of operands $\Phi$ and
$\Psi$ as well as the attached by the conjunction $\&$ the binary
conditional $\theta$. The equation in a neutral form is as shown
in Figure~\ref{fig-06}. For instance, assume that $\theta \in \{=,
\neq, <, >, \le \ge\}$, and establish an evaluation procedure as
follows:
\begin{description}
\item select out the needed join operation $\theta$;
\item select out the operands $\Phi$ and $\Psi$;
\item select out the elements $h$ and $\hbar$ from the corresponding
operands;
\item the result of join is generated by selecting out
      and accumulating the evaluated data objects $h$ and $\hbar$
      which match the formula $\theta([h,\hbar])$.
\end{description}

\subparagraph{Statics}

Let evolvent $f$ be the {\em identity} map $f\equiv 1_I : I \to
I$. The most significant case is for occurrences of data objects
$\overline{\hbar}$ within the formulae
$\theta[\overline{h},\overline{\hbar}]$:
$$
\theta[\overline{h}, \left \{ \begin{array}{l}
                                 \overline{x}    \\
                                 g(\overline{x}) \\
               \lbrack   \overline{x},\overline{y} \rbrack \\
                                 \overline{x}(\overline{y})
                                   \end{array}
                     \right \} ]
$$
For a constant binary relation $\theta$ and a constant function
$g$ the evaluation depends on the cases of $\overline{\hbar}$:
\[
\begin{array}{llcl}
\textrm{atomic object}
&\|\theta[\overline{h},\overline{x}]\| &=&
        \theta \circ \langle h, x \rangle \\

\textrm{constant function}
&\|\theta[\overline{h},g(\overline{x})]\| &=&
        \theta \circ \langle h, g \circ x \rangle \\
\textrm{ordered pair}
&\|\theta[\overline{h},[\overline{x},\overline{y}]]\| &=&
        \theta \circ \langle h,
         \langle x, y \rangle \rangle  \\
\textrm{application}
&\|\theta[\overline{h},\overline{x}(\overline{y})]\| &=&
        \theta \circ \langle h,
         \varepsilon \circ \langle x_{1_I}, y \rangle \rangle
\end{array}
\]

\subparagraph{Dynamics}

The dynamical behavior with evolvent $f : B \to I$ drops down to
the equations with atomic objects depending on the same cases of
$\overline{\hbar}$ where $i\in I$ and $b\in B$ as shown in
Figure~\ref{fig-07}.
Note that $i$ and $b$ have a meaning of the
{\em views}.

\section*{Conclusions}
\addcontentsline{toc}{section}{Conclusions} Some topics concerning
the usefulness of partial element are briefly outlined. At first,
the partial elements are used for generating the variable domains,
and this is a generalization of types giving rise to their valid
families depending on the sequences of events.

Second, the partial elements as data objects are determined
using the formal descriptions. The approach leads to
generating the additional term. The class of terms as was shown
contains the restricted relations which are formula driven.

Third, the partial elements are applied to study the basic
operation within event sensitive query language.
\addcontentsline{toc}{section}{References}
{\small

\begin{thebibliography}{EGS91}

\bibitem[Bee90]{Beer:90}
C.~Beeri.
\newblock A formal approach to object oriented databases.
\newblock {\em Data{\&}Knowledge Engineering}, 5:353--382, 1990.

\bibitem[CW85]{CaWe:85}
L.~Cardelli and P.~Wegner.
\newblock On understanding types, data abstractions, and polymorphism.
\newblock {\em Computing Syrveys}, 17(4), December 1985.

\bibitem[EGS91]{EhGoSe:91}
H.-D. Ehrich, M.~Gogola, and A.~Sernadas.
\newblock A categorial theory of objects as observed processes.
\newblock In J.W. {d}e{B}akker~et. al., editor, {\em {P}roceedings of the
  {REX/FOOL} {S}chool/{W}orkshop}, volume 489 of {\em {L}ecture {N}otes in
  {C}omputer {S}cience}, pages 203--228. Berlin, Heidelberg, New York, Springer
  Verlag, 1991.

\bibitem[Ism98]{unpublished-LYuI:98a}
L.Yu. Ismailova.
\newblock Event driven computations for relational model.
\newblock Talk at JMSU Institute for Contemporary Education, March, April 1998.

\bibitem[IZ96]{LYuIZ:96}
L.~Yu. Ismailova and K.E. Zinchenko.
\newblock Object-oriented tools for advanced applications.
\newblock In B.~Novikov and J.W. Schmidt, editors, {\em Proceedings of the 3rd
  International Workshop on Advances in Databases and Information Systems},
  volume~2, pages 27--31, Moscow, September 1996. Moscow Engineering Physical
  Institute, Moscow 1996.

\bibitem[IZ97]{LYuIZ:97}
L.~Yu. Ismailova and K.E. Zinchenko.
\newblock An object evaluator to generate flexible applications.
\newblock In V.~Wolfengagen and R.~Manthey, editors, {\em Proceedings of the
  1st East-European Symposium on Advances in Databases and Information
  Systems}, volume~1, pages 141--148, St.-Petersburg, September 1997. Nevsky
  Dialect, St.-Petersburg 1997.

\bibitem[MB90]{Mano:90}
F.~Manola and A.P. Buchmann.
\newblock A functional/relational object-oriented model for distributed data
  management: Preliminary description.
\newblock {TM}-0331-11-90-165, GTE Laboratories Incorporated, December 31 1990.

\bibitem[Sco71]{Sco:71}
D.S. Scott.
\newblock The lattice of flow diagrams.
\newblock In {\em Symposium on semantics of algorithmic languages}, volume 188
  of {\em Lecture notes in mathematics}, pages 311--378. Berlin, Heidelberg,
  New York, Springer Verlag, 1971.

\bibitem[Sco80]{Sco:80}
D.S. Scott.
\newblock Relating theories of the $\lambda$-calculus.
\newblock In J.~Hinhley and J.~Seldin, editors, {\em To H.B. Curry: Essays on
  combinatory logic, lambda calculus and formalism}, pages 403--450. New York
  and London, Academic Press, 1980.

\bibitem[Wol98]{unpublished-Wol:98a}
V.E. Wolfengagen.
\newblock Event driven objects: {P}art {I}.
\newblock Talk at JMSU Institute for Contemporary Education, March, April 1998.

\end{thebibliography}
\newcommand{\noopsort}[1]{} \newcommand{\printfirst}[2]{#1}
  \newcommand{\singleletter}[1]{#1} \newcommand{\switchargs}[2]{#2#1}

}

\end{document}